\documentclass[conference]{IEEEtran}
\IEEEoverridecommandlockouts
\usepackage{cite}
\usepackage{amsmath,amssymb,amsfonts,latexsym}
\usepackage{algorithmic}
\usepackage{graphicx}
\usepackage{subcaption}
\usepackage{textcomp}
\usepackage{xcolor}
\def\BibTeX{{\rm B\kern-.05em{\sc i\kern-.025em b}\kern-.08em
    T\kern-.1667em\lower.7ex\hbox{E}\kern-.125emX}}

\usepackage{mathtools}
\usepackage{comment}
\usepackage{hyperref}
\usepackage{booktabs} 
\usepackage{url}            

\ifCLASSINFOpdf
\else
\fi

\usepackage{amssymb}
\usepackage{amsmath}
\usepackage{titlesec}
\usepackage{url}
\usepackage{makecell}
\usepackage{xcolor}

\begin{document}
\title{Resting-state fMRI Analysis using Quantum Time-series Transformer \thanks{The views expressed in this article are those of the authors and do not represent the views of Wells Fargo. This article is for informational purposes only. Nothing contained in this article should be construed as investment advice. Wells Fargo makes no express or implied warranties and expressly disclaims all legal, tax, and accounting implications related to this article.}}

\author{Junghoon Justin Park$^1$, Jungwoo Seo$^2$, Sangyoon Bae$^1$, Samuel Yen-Chi Chen$^3$, Huan-Hsin Tseng$^4$ \\ Jiook Cha$^{1,2,5}$, Shinjae Yoo$^4$\\
\small $^1$Interdisciplinary Program in Artificial Intelligence, Seoul National University \\ $^2$Department of Brain and Cognitive Sciences, Seoul National University\\ $^3$Wells Fargo \\ $^4$Artificial Intelligence Department, Brookhaven National Laboratory \\ $^5$Department of Psychology, Seoul National University \\
\small \texttt{utopie9090@snu.ac.kr, jungwoo.seo95@gmail.com, stellasybae@snu.ac.kr,} \\
\small \texttt{yen-chi.chen@wellsfargo.com, htseng@bnl.gov, connectome@snu.ac.kr, sjyoo@bnl.gov}
}


\maketitle

\begin{abstract}
Resting-state functional magnetic resonance imaging (fMRI) has emerged as a pivotal tool for revealing intrinsic brain network connectivity and identifying neural biomarkers of neuropsychiatric conditions. However, classical self-attention transformer models—despite their formidable representational power—struggle with quadratic complexity, large parameter counts, and substantial data requirements. To address these barriers, we introduce a Quantum Time-series Transformer, a novel quantum-enhanced transformer architecture leveraging Linear Combination of Unitaries and Quantum Singular Value Transformation. Unlike classical transformers, Quantum Time-series Transformer operates with polylogarithmic computational complexity, markedly reducing training overhead and enabling robust performance even with fewer parameters and limited sample sizes. Empirical evaluation on the largest-scale fMRI datasets from the Adolescent Brain Cognitive Development Study and the UK Biobank demonstrates that Quantum Time-series Transformer achieves comparable or superior predictive performance compared to state-of-the-art classical transformer models, with especially pronounced gains in small-sample scenarios. Interpretability analyses using SHapley Additive exPlanations further reveal that Quantum Time-series Transformer reliably identifies clinically meaningful neural biomarkers of attention-deficit/hyperactivity disorder (ADHD). These findings underscore the promise of quantum-enhanced transformers in advancing computational neuroscience by more efficiently modeling complex spatio-temporal dynamics and improving clinical interpretability.
\end{abstract}


%
\IEEEpeerreviewmaketitle

\section{Introduction}
Resting-state functional magnetic resonance imaging (fMRI) has emerged as a critical tool in neuroscience and psychiatry, significantly advancing our understanding of brain function, connectivity, and the underlying mechanisms of various neurological and psychiatric disorders \cite{Lv, Lee2018, Gonzalez-Castillo}. The scientific and clinical significance of resting-state fMRI lies in its ability to reveal intrinsic connectivity patterns within the brain. These connectivity patterns, representing synchronized neuronal activities across brain regions, have been associated with various cognitive functions and pathological states, including attention-deficit/hyperactivity disorder (ADHD) \cite{Wang2023}, psychosis \cite{Karcher}, major depressive disorder \cite{Bondi}, and other neuropsychiatric conditions \cite{Smith, Marek2019}. Detecting altered connectivity patterns has clinical implications, enabling earlier diagnosis, prognosis assessment, and targeted therapeutic strategies.

Recent advancements in machine learning and deep learning techniques, particularly spatio-temporal learning algorithms such as Transformers \cite{Vaswani}, have provided powerful new tools for analyzing complex resting-state fMRI data \cite{Asadi, BNT, BolT}. Transformers, characterized by self-attention mechanisms, effectively model dynamic interactions across spatial and temporal dimensions, capturing intricate brain network dynamics better than traditional correlation-based approaches. These methods have demonstrated superior performance in decoding clinical and cognitive outcomes from resting-state fMRI data, highlighting their potential to enhance precision medicine applications in psychiatry.

Despite their advantages, classical transformer-based algorithms have notable limitations. Transformers typically require segmenting fMRI data into fixed-size temporal windows, which can lead to potential loss of information regarding long-term temporal dynamics and dependencies \cite{Asadi}. Standard positional embedding methods may inadequately capture complex temporal dependencies inherent in neural time-series data, restricting transformers' capacity to model intricate spatiotemporal relationships \cite{Vaswani}. Moreover, classical transformer models involve millions of parameters, necessitating large-scale datasets for training. However, datasets of such magnitude are not always available in neuroimaging research, limiting their applicability \cite{Marek2022}. Additionally, training transformers with millions of parameters demands extensive computational resources, including high-performance GPUs, which impose high energy consumption and carbon emissions, raising environmental sustainability concerns \cite{Patterson}.

To address these limitations, this study introduces a Quantum Time-series Transformer algorithm designed to analyze resting-state fMRI data by leveraging quantum computing principles such as quantum entanglement and superposition. By employing these quantum mechanical concepts, the Quantum Time-series Transformer algorithm can inherently capture complex spatio-temporal dependencies more effectively than classical positional embedding approaches. Additionally, the quantum architecture offers improved computational efficiency, enhanced representational power, and better sustainability through reduced parameterization and computational resource demands. Consequently, the proposed Quantum Time-series Transformer method provides a promising alternative, particularly suited to the typically smaller sample sizes characteristic of neuroimaging research, thereby offering significant advantages over classical transformer-based models.

\section{Background}
\subsection{Variational quantum circuits}
Variational quantum circuits (VQCs), also known as parameterized quantum circuits or quantum neural networks, constitute a foundational paradigm within quantum machine learning (QML), bridging classical computational tasks and quantum computation. Typically, a VQC comprises three essential components: an encoding circuit, a parameterized ansatz circuit, and a quantum measurement stage.

The encoding circuit, denoted as $U(\overrightarrow{x})$, is specifically designed to encode a classical input vector $\overrightarrow{x}$ into a quantum state $U(\overrightarrow{x})|0\rangle^{\otimes n}$, where $|0\rangle^{\otimes n}$ represents the initial ground state for an $n$-qubit quantum system. This process effectively maps classical data into the exponentially large Hilbert space of quantum states, enabling potentially significant quantum advantage.

Subsequently, the encoded quantum state undergoes transformation through the parameterized ansatz circuit $W(\Theta)$, composed of multiple trainable quantum gate layers. This results in the state:
\begin{equation}
    |\psi(\overrightarrow{x}, \Theta)\rangle = W(\Theta)U(\overrightarrow{x})|0\rangle^{\otimes n} = \left(\prod_{j=M}^{1} V_j(\overrightarrow{\theta_j})\right)U(\overrightarrow{x})|0\rangle^{\otimes n},
\label{QuantumState}
\end{equation}
where each $V_j(\overrightarrow{\theta_j})$ represents a layer of quantum gates parameterized by the vector $\overrightarrow{\theta_j}$, and $\Theta=\{\overrightarrow{\theta_1}, \ldots, \overrightarrow{\theta_M}\}$ collectively denotes the set of all adjustable parameters. Each parameterized gate layer typically comprises rotations and entanglement gates, enabling intricate manipulations of quantum state distributions.

Information embedded in the resulting quantum state is extracted via quantum measurements defined by specific quantum observables $\hat{H}_k$. Typically, these observables are Hermitian operators, chosen for their property of yielding real-valued expectation values, critical for classical interpretation and subsequent optimization. Thus, a VQC implements a quantum parametric function:
\begin{equation}
\overrightarrow{f}(\overrightarrow{x}; \Theta) = \left(\langle \hat{H}_1 \rangle, \ldots, \langle \hat{H}_n \rangle\right),
\end{equation}
with each expectation value calculated as:
\begin{equation}
\langle \hat{H}_k \rangle = \langle 0|U^{\dagger}(\overrightarrow{x})W^{\dagger}(\Theta)\hat{H}_k W(\Theta)U(\overrightarrow{x})|0\rangle.
\end{equation}
Practically, these expectation values are estimated through repeated measurements (shots) on quantum hardware, or via precise computation on quantum simulation platforms.

During training, the parameters $\Theta$ are iteratively adjusted to optimize the performance of the quantum circuit $f_{\Theta}(\overrightarrow{x})$ by minimizing a predefined loss function $\mathcal{L}(\overrightarrow{x}, y; \Theta)$ over the training dataset $D=\{(\overrightarrow{x}_i, y_i)\}_i$. For instance, in a regression scenario, the loss function typically takes the form:
\begin{equation}
\mathcal{L}(\overrightarrow{x}, y; \Theta) = \sum_i \| f_{\Theta}(\overrightarrow{x}_i) - y_i \|^2.
\end{equation}
The values of the loss $\mathcal{L}(\overrightarrow{x},y; \Theta)$ (or its gradient) are estimated from the quantum circuit outputs and subsequently provided to a classical optimization algorithm, which attempts to solve the optimization problem:
\begin{equation}
\arg\min_{\Theta}\mathcal{L}(\overrightarrow{x},y; \Theta).
\end{equation}

A Hermitian operator $H$, fundamental in quantum mechanics, satisfies the condition $H=H^\dagger$. Such operators have the property that expectation values $\langle\psi|H|\psi\rangle$ are always real-valued. An $N\times N$ Hermitian matrix ($N=2^n$ for an $n$-qubit system) contains $N^2$ real parameters and can be explicitly represented as:
\begin{equation}
    H = \sum_{i,j=1}^{N} h_{ij}E_{ij}, \quad \text{with } h_{ij}=\overline{h_{ji}},
\end{equation}
where $E_{ij}$ is the elementary matrix with exactly one nonzero entry located at the $(i,j)$ position:
\begin{equation}
    (E_{ij})_{kl}=\delta_{ik}\delta_{jl}.
\end{equation}

Given a quantum state $|\psi\rangle$, such as the state defined in Equation \ref{QuantumState}, the expectation value with respect to a parameterized Hermitian operator $H(\overrightarrow{h})$ is:
\begin{equation}
    \langle\psi|H(\overrightarrow{h})|\psi\rangle = \sum_{i,j=1}^{N}h_{ij}\langle\psi|E_{ij}|\psi\rangle.
\end{equation}
This explicit form enables efficient gradient-based optimization. Specifically, the gradient of the expectation with respect to parameters $\overrightarrow{h}$ can be computed as:
\begin{align}
    \frac{\partial \langle\psi|H(\overrightarrow{h})|\psi\rangle}{\partial h_{kl}} &= \langle\psi|E_{kl}|\psi\rangle \nonumber \\
    &= \left(\overline{W(\Theta)U(\overrightarrow{x})}\right)_{k1}\left(W(\Theta)U(\overrightarrow{x})\right)_{l1},
\end{align}
highlighting a clear dependence on both the parameterized quantum circuit $W(\Theta)$ and the encoded quantum representation of input data.

In summary, VQCs offer a robust and adaptable framework capable of leveraging quantum mechanical phenomena, such as superposition and entanglement, to enhance computational performance in machine learning tasks, thereby establishing their critical role in emerging quantum technologies.

\subsection{Classical Self-Attention Transformer}
Classical transformers have significantly advanced deep learning through their powerful self-attention mechanism, effectively capturing long-range dependencies in sequential data \cite{Vaswani}. Self-attention enables transformers to directly model interactions between all pairs of input tokens, resulting in dynamic and contextually rich representations.

\subsubsection{Operation of Self-Attention Transformers}
Self-attention transformers process input sequences through three main steps: query, key, and value computation. First, input tokens are linearly projected into three distinct vectors: queries (Q), keys (K), and values (V). The self-attention mechanism then computes attention scores using scaled dot-product operations between queries and keys:
\begin{equation}
    Attention(Q,K,V)=softmax(\frac{QK^T}{\sqrt{d_k}})V.
\end{equation}
Here, $Q, K$, and $V$ represent query, key, and value matrices of sizes $L\times d$, respectively, each derived by learned linear transformations of the input embeddings. The matrix operation $QK^T$ produces an attention matrix of size $L \times L$ and the softmax operation generates weights that quantify the relevance of each token relative to others in the input sequence. The model applies multiple parallel attention "heads," enabling the representation of different contextual aspects and enhancing expressive power.

A transformer consists of stacked layers, each composed of self-attention followed by a feed-forward neural network. Residual connections and layer normalization stabilize training and improve learning efficiency.

\subsubsection{Limitations of Classical Self-Attention Transformers in fMRI Analysis}
Despite their success in language modeling, classical self-attention transformers have several notable limitations when applied to resting-state fMRI data:

\paragraph{Fixed-Size Temporal Window Segmentation}
Transformers typically process inputs segmented into fixed-size temporal windows. While convenient computationally, this approach restricts their ability to capture continuous, long-term temporal dynamics and dependencies inherent in neural time-series data. Such segmentation can inadvertently omit essential temporal context spanning multiple segments, impairing model performance.

\paragraph{Positional Embedding Constraints}
Standard positional embedding methods (e.g., sinusoidal or learned positional embeddings) commonly used in classical transformers may inadequately represent complex temporal dependencies and nonlinearities in brain activity data. fMRI signals exhibit sophisticated temporal dynamics and multifractal properties, posing a significant challenge for simple positional embedding schemes, limiting transformers' ability to accurately capture intricate spatio-temporal relationships.

\paragraph{Quadratic Computational Complexity}
Classical transformers exhibit quadratic computational complexity ($\mathcal{O}(L^2)$), where $L$ is the sequence length. The computational bottleneck is the $L\times L$ matrix multiplication for computing attention scores $QK^T$. This quadratic complexity significantly increases computational demands for both training and inference, necessitating extensive computational resources and high-performance GPUs. Consequently, these resource-intensive requirements result in substantial energy consumption, associated with considerable environmental impacts, including high electricity usage and carbon emissions.

\paragraph{Limited Availability of Large-scale Neuroimaging Datasets}
Transformers typically require extensive datasets for training due to their vast number of parameters. However, neuroscience and psychiatry research often deal with relatively small sample sizes—frequently fewer than a hundred participants. Such limited datasets are insufficient for effectively training classical transformer models, thereby severely constraining their applicability in resting-state fMRI analysis. Additionally, the small sample sizes commonly found in neuroimaging research exacerbate problems of overfitting and limit the generalizability, reproducibility, and interpretability of transformer-based findings \cite{Marek2022, Yarkoni, Schnack}.

These limitations highlight critical opportunities for innovative approaches, such as quantum transformers, to address both representational and computational efficiency challenges associated with classical transformer models in resting-state fMRI analysis.

\subsection{Linear Combination of Unitaries and Quantum Singular Value Transformation}
Recent advancements in classical machine learning, including transformer-based architectures, typically involve large-scale matrix computations to achieve powerful representational capabilities \cite{Hwang}. These classical algorithms predominantly perform linear and nonlinear transformations through matrix operations on input embeddings. With advancements in quantum computing, quantum analogs of these classical matrix operations have become feasible through techniques like the Linear Combination of Unitaries (LCU) and Quantum Singular Value Transformation (QSVT). Leveraging these quantum algorithmic primitives, classical machine learning models such as transformers can be effectively adapted into QML algorithms.

\subsubsection{Linear Combination of Unitaries}
Linear Combination of Unitaries (LCU) is a powerful quantum algorithmic primitive used to implement complex linear operations as a weighted sum of simpler quantum unitary operations \cite{Childs}. Given a set of unitary operations $\{U_j\}$ and complex coefficients $\{b_j\}$, LCU allows for constructing and executing a linear combination:
\begin{equation}
    M=\sum_jb_jU_j, \quad s.t. \quad \sum_j|b_j| \leq 1.
\end{equation}
To implement $M$ on a quantum state, an ancillary quantum register is introduced and initialized to a superposition state weighted by $\{b_j\}$. Controlled operations then apply the unitaries $U_j$ conditioned on the ancillary register state, resulting in the desired combination after successful postselection. Thus, classical matrix operations fundamental to transformers, like linear mixing of embeddings, can be directly implemented using LCU.

\subsubsection{Quantum Singular Value Transformation}
Quantum Singular Value Transformation is an advanced quantum primitive enabling nonlinear transformations of matrices encoded within quantum circuits \cite{Gilyen}. QSVT applies polynomial transformations to the singular values of an input matrix $M$ by employing parameterized quantum gates. Given a polynomial $P_{\overrightarrow{c}}(x)$, defined as:
\begin{equation}
    P_{\overrightarrow{c}}(x) = c_dx^d+c_{d-1}x^{d-1}+ \cdots+c_1x+c_0,
\end{equation}
the QSVT circuit applies this polynomial transformation to a block-encoded matrix $M$, resulting in:
\begin{equation}
    P_{\overrightarrow{c}}(M) = c_dM^d+c_{d-1}M^{d-1}+ \cdots+c_1M+c_0I.
\end{equation}
Here, the transformation is performed by repeatedly applying controlled unitary and controlled adjoint unitary operations interleaved with specific phase shifts to encode polynomial coefficients. QSVT inherently introduces nonlinearity to quantum circuits, analogous to the nonlinear transformations applied by activation functions in classical neural networks.

\subsubsection{Conversion of Classical Matrix Operations to Quantum Algorithms}
By integrating LCU and QSVT, classical machine learning algorithms based on linear and nonlinear matrix operations can be systematically converted into quantum algorithms. Classical transformer operations—such as self-attention mechanisms, linear transformations, and nonlinear activation functions—can be restructured into a sequence of unitary transformations and polynomial manipulations enabled by LCU and QSVT. Such quantum transformations potentially offer enhanced computational efficiency due to inherent parallelism, reduced memory requirements, and fundamentally different ways of embedding and processing information, making them especially promising for analyzing complex spatio-temporal data such as resting-state fMRI.

\section{Quantum Time-series Transformer}
The Quantum Time-series Transformer presented in this study adapts the principles of the recently proposed quantum transformer architecture, \textit{Quixer} \cite{Quixer}, for effective representation and analysis of spatio-temporal sequences inherent in resting-state fMRI data. Unlike classical transformers, which employ matrix operations on classical embeddings, the Quantum Time-series Transformer leverages quantum mechanical principles—such as superposition and entanglement—to encode and process spatio-temporal information within quantum states.

\begin{figure*}[h]
    \centering
    \includegraphics[width = 0.9\textwidth]{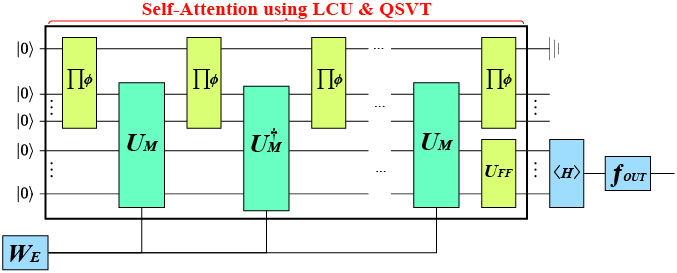}
    \caption{Model Architecture of the Quantum Time-series Transformer. The model begins by linearly mapping temporal embeddings to angles using the matrix $W_E$. These angles parameterize unitary quantum circuits that act on a data register. A linear superposition of these unitaries is generated using an LCU operator $U_M$, which functions similarly to the self-attention mechanism in classical transformers. It prepares a superposition of unitary operations that allows the model to dynamically adjust how different temporal elements are processed, akin to the way self-attention weights different parts of a sequence. Following this, a polynomial transformation is applied through the phases in the $\Pi_\phi$ gates, using the QSVT procedure. This transformation further adjusts the processing, allowing the model to capture complex temporal dependencies within the data. Next, a feed-forward unitary $U_{FF}$ is applied to the data register. The data is then measured from the resulting quantum state using multiple measurement operators, and the expectation values $\langle H \rangle$ are classically processed by $f_{out}$ to produce the final output of the model. Adapted from the Quixer model architecture in Khatri et al. \cite{Quixer}} 
    \label{Fig_QuantumTSTransformer}
\end{figure*}

\subsection{How the Quantum Time-series Transformer Works}
At its core, the Quantum Time-Series Transformer processes input data through three main quantum steps: quantum sequence embedding, quantum mixing, and quantum nonlinear transformation.

\subsubsection{Quantum Sequence Embedding}
The Quantum Time-series Transformer begins by converting classical spatio-temporal data (e.g., signals from brain regions across time) into quantum embeddings. Specifically, classical vector embeddings representing individual spatio-temporal sequences are linearly transformed into angular parameters, which are then used to parameterize quantum gates within VQCs. Each sequence thus obtains a unique quantum circuit representation, facilitating quantum-based computation.

\subsubsection{Quantum Mixing via LCU}
After quantum embedding, each spatio-temporal sequence is represented by a corresponding unitary operation. These unitary operations are combined using the LCU method, creating a weighted superposition of quantum embeddings. This approach inherently leverages quantum superposition, enabling simultaneous representation and interaction of multiple sequences, thereby enhancing the model’s capability to capture complex spatio-temporal dynamics.

\subsubsection{Nonlinearity through QSVT}
Following the linear combination step, QSVT introduces nonlinearity into the quantum embeddings by applying polynomial transformations. These polynomial transformations, as detailed previously, allow the Quantum Transformer to effectively model complex, nonlinear interactions among spatio-temporal sequences. Such nonlinear transformations significantly enhance the model’s ability to represent the intricate dependencies observed in resting-state fMRI data.

\subsubsection{Quantum Attention via Unitary Composition}
Classical transformers explicitly compute pairwise interactions among sequences through attention mechanisms. By contrast, the Quantum Time-series Transformer naturally captures these interactions through quantum mechanical principles such as entanglement, realized by composing sequence representations as quantum states. Through this unitary composition, the model implicitly encodes attention-like behavior, effectively representing rich, nuanced interactions and spatio-temporal dependencies without explicitly calculating all pairwise attention scores.

\subsubsection{Final State Measurement and Output Processing}
After applying quantum transformations, the resultant quantum state is measured using quantum observables—such as expectation values of Pauli operators. The measured values, reflecting quantum state information, are processed using a classical feed-forward neural network to generate the final model output, typically used for classification or regression tasks relevant to neuroscience research.

\subsection{Advantages of Quantum Time-series Transformer}
The Quantum Time-series Transformer, by integrating quantum computational primitives such as the LCU and QSVT, offers significant advantages over classical transformer algorithms in analyzing resting-state fMRI data.

\subsubsection{Improved Spatio-Temporal Representation}
Classical transformer models rely primarily on linear operations combined with simplistic positional embeddings, making it challenging to fully capture the complex, nonlinear, and hierarchical dynamics inherent in spatio-temporal brain data. By contrast, the Quantum Time-series Transformer leverages fundamental quantum mechanics principles, including quantum entanglement and superposition, to naturally and effectively represent complex interdependencies and nonlinear interactions within brain data. Quantum embedding through VQCs encodes data into quantum states, inherently capturing intricate spatial and temporal relationships that may be difficult or inefficient for classical embeddings to model.

Moreover, quantum state representations allow for the simultaneous encoding of global context through entanglement, facilitating the capture of long-range and multiscale dependencies without explicit positional embedding methods. This inherent global contextualization significantly enhances representational power and flexibility compared to classical transformer architectures.

\subsubsection{Enhanced Computational Efficiency}
The Quantum Time-series Transformer, leveraging quantum computational primitives such as LCU and QSVT, significantly enhances computational efficiency compared to classical self-attention transformer algorithms. Classical self-attention transformers suffer from quadratic computational complexity, primarily due to the explicit calculation and storage of the attention matrix, which involves pairwise computations between all sequence elements \cite{Vaswani}. Specifically, classical self-attention complexity scales quadratically with the sequence length $L$, represented as $\mathcal{O}(L^2d)$, where $d$ is the embedding dimension.

In contrast, the Quantum Time-series Transformer leverages quantum primitives—LCU and QSVT—to fundamentally alter the computational paradigm. These quantum primitives do not explicitly compute or store the complete attention matrix. Instead, LCU prepares quantum states representing linear combinations of sequence embeddings in a compact, quantum-mechanical form, effectively avoiding the explicit computation of the entire $L \times L$ attention matrix. Subsequently, QSVT applies polynomial-based nonlinear transformations on these quantum states, efficiently emulating the nonlinear computations performed by classical transformers, such as the softmax function.

The computational complexity of quantum transformers using the combined LCU and QSVT procedures scales polylogarithmically with sequence length, represented as $\mathcal{O}(polylog(L))$. Thus, the quantum complexity comparison with classical algorithms is as follows:
\begin{table}[h]
    \centering
    \caption{Comparison of Computational Complexity}    
    \begin{tabular}{ c c c }
    \toprule
 \textbf{Operation}& \textbf{\makecell{Classical\\ Complexity}}&\textbf{\makecell{Quantum \\Complexity}}\\
     \toprule
         Computing Attention Scores& $\mathcal{O}(L^2d)$ & $\mathcal{O}(polylog(L))$\\
         Applying Nonlinear Transformations & $\mathcal{O}(L^2)$ & $\mathcal{O}(polylog(L))$\\
          Application to Value matrix (V) & $\mathcal{O}(L^2d)$ & $\mathcal{O}(polylog(L))$\\
\bottomrule
    \end{tabular}    
\label{Complexity_table}
\end{table}

By reducing complexity from quadratic to polylogarithmic scaling (Table \ref{Complexity_table}), the Quantum Time-series Transformer not only significantly enhances computational efficiency but also allows for effective modeling with fewer computational resources, lower energy consumption, and improved sustainability, addressing critical concerns related to energy usage and carbon emissions associated with large-scale classical transformer training \cite{Jaschke}.

\subsubsection{Improved Generalizability with Small Sample Sizes}
A fundamental challenge in neuroimaging research is the limited availability of large-scale datasets \cite{Marek2022}. Classical transformer models typically require extensive training datasets to perform effectively due to their high parameterization. In contrast, Quantum Time-series Transformer can achieve high generalization performance even with a small number of parameters and limited training data.

The ultimate goal of a QML model is to minimize the expected loss $R(\Theta)$ over the data distribution $P$, formally expressed as:
\begin{equation}
R(\Theta) = \mathbb{E}_{\overrightarrow{x}\sim P}[\mathcal{L}(\overrightarrow{x}, y;\Theta)].
\end{equation}
Given that the true distribution $P$ is unknown, $R(\Theta)$ is approximated using a finite training set $S=\{\overrightarrow{x_i}, y_i\}_{i=1}^{N}$, yielding the training loss:
\begin{equation}
\hat{R}_S(\Theta) = \frac{1}{N}\sum_{i=1}^{N}\mathcal{L}(\overrightarrow{x}_i, y_i;\Theta).
\end{equation}
The generalization error, defined as:
\begin{equation}
gen(\Theta) = R(\Theta)-\hat{R}_S(\Theta),
\end{equation}
quantifies the ability of the QML model to perform effectively on unseen data \cite{Caro2021}.

Recent studies highlight that VQCs can represent matrix product states with exponentially large bond dimensions, reflecting a distinct quantum advantage in representational capacity compared to classical neural networks \cite{Du2020}. This superior expressibility primarily arises from quantum properties such as entanglement and superposition, enabling VQCs to efficiently represent complex, high-dimensional quantum states and distributions otherwise inaccessible or inefficiently approximated by classical methods \cite{Du2020, Zhao2024}.

The use of quantum entanglement and superposition within parameterized quantum gates allows VQCs to achieve faster training, higher effective dimensionality, and broader coverage of the function space, resulting in enhanced generalization performance \cite{Abbas}. Specifically, increased quantum entanglement within a circuit tends to reduce both the expected loss $R(\Theta)$ and the training loss $\hat{R}_S(\Theta)$, thereby decreasing the generalization error, even when utilizing relatively small training datasets and limited parameters \cite{Caro2022}.

Formally, the generalization error $gen(\Theta)$ for QML models composed of $T$ parameterized local quantum channels has been theoretically characterized by:
\begin{equation}
gen(\Theta^{*}) \in \mathcal{O}\left(\sqrt{\frac{T\log T}{N}}\right),
\end{equation}
where $N$ represents the training data size \cite{Caro2022}. This relationship underscores the inherent efficiency and generalizability advantages of QML models, especially in scenarios with sparse or limited datasets.

Therefore, Quantum Time-series Transformers can effectively leverage the representational efficiency and expressiveness provided by quantum states, enabling robust performance even with the relatively small sample sizes typical in neuroscience studies. Consequently, quantum models potentially exhibit enhanced generalizability and reduced risk of overfitting, significantly improving the reliability and reproducibility of analyses in neuroscience and psychiatry research.

\section{Experiment}
 We trained our Quantum Time-series Transformer on two large-scale resting-state fMRI datasets: the Adolescent Brain Cognitive Development (ABCD) Study and the UK Biobank (UKB). The ABCD study tracks brain development in 11,868 children aged 9-10 years, offering a rich dataset for analyzing age-related and psychiatric disorder-related brain network dynamics through resting-state fMRI \cite{Casey}. The UK Biobank consists of brain imaging and extensive phenotypic data from 41,283 participants aged 40-77 years, enabling the study of brain structure, function, and their associations with various health outcomes in a population-based context \cite{Miller}. For both datasets, we used the HCP-MMP1 ROI atlas \cite{Glasser} for brain region definition. This multi-modal cortical parcellation combines structural, functional, and connectivity measures, providing high-resolution, data-driven cortical area mappings that are crucial for precise neuroanatomical and functional connectivity analyses.

The objective of our model was to process and learn the spatio-temporal dynamics of resting-state fMRI data to predict various phenotypes. The phenotypes of interest were biological sex and ADHD diagnosis (binary classification), and fluid intelligence (regression). ADHD diagnosis was based on CBCL attention and ADHD scores exceeding a T-score threshold of 65, with healthy controls defined as individuals without any diagnosed mental health disorders \cite{Cordova}. This classification scheme ensured a clear separation between clinical and non-clinical groups, enhancing model training and evaluation.

We first performed a full sample analysis for each task phenotype, using samples that underwent quality control and had no missing observations. To assess the model's generalizability with limited training data, we conducted a small sample analysis using 100 randomly selected participants from the full sample. Both full and small sample analyses were repeated for multiple random seeds to ensure robustness. Importantly, the random selection of 100 participants for each small sample analysis was independent for each seed, providing a valid assessment of generalizability.

For comparison, we evaluated the performance of the Quantum Time-series Transformer against three classical transformer models: the vanilla transformer \cite{Vaswani}, the Brain Network Transformer, and BolT. The Brain Network Transformer models brain networks as graphs, using connection profiles as node features to capture functional connectivity in resting-state fMRI data \cite{BNT}. Its self-attention mechanism learns pairwise interaction strengths among regions of interest (ROIs), making it well-suited for brain network analysis in neuroimaging tasks. These models represent established transformer architectures in the classical machine learning domain. BolT is a transformer model designed for fMRI time-series analysis, utilizing a fused window attention mechanism to efficiently capture both local and global temporal representations \cite{BolT}. It improves upon vanilla transformers by using overlapping temporal windows and cross-window attention, which is essential for analyzing dynamic fMRI data. These two models were chosen as baselines for comparison due to their state-of-the-art performance in brain network and time-series fMRI analysis, providing a solid comparison against the quantum-powered Quantum Time-series Transformer in our study.

To ensure a fair comparison, all classical transformer models were designed to match the Quantum Time-series Transformer as closely as possible. The encoding and post-processing layers were identical across all quantum and classical models, with differences confined to the self-attention mechanisms, which vary between the models.

SHAP values \cite{Lundberg} are computed to interpret the contributions of individual features to the model’s predictions. In the context of the time-series quantum transformer model, SHapley Additive exPlanations (SHAP) values are derived by evaluating the difference in the model's output when a feature is occluded (replaced with a baseline value) versus when it is included in the input. Specifically, for each feature, the SHAP value is calculated as the difference in prediction between the full input and the input with the feature set to its baseline value, preserving the sign of the prediction. This approach allows us to measure the marginal contribution of each feature to the model’s output. For each instance, the SHAP values represent the individual impact of time-series features, enabling a detailed interpretation of how specific features, such as different brain regions or temporal patterns, influence the model's decision. This method provides an intuitive, data-driven explanation of the complex dynamics in the time-series data and helps elucidate the role of each feature in the quantum model's decision-making process \cite{Heese, Lundberg, Burge, Pira}.

All experiments utilized the TorchQuantum library \cite{TorchQuantum}, integrated with PyTorch for seamless quantum-classical hybrid computations. The experiments were conducted on a Linux server (Kernel 5.14) with 128 CPU cores, 256 threads (x86-64 architecture), 503.14 GB RAM, and an NVIDIA A100-PCIE GPU with 40 GB memory. The software environment included Python 3.11.7, PyTorch 2.5.0+cu121, and CUDA 12.1.

\section{Results}

In the full sample analysis, the Quantum Time-series Transformer outperformed classical models in several key metrics. It achieved the lowest mean absolute error (MAE) in predicting fluid intelligence for the ABCD dataset and the highest area under the receiver operating characteristic curve (AUROC) for biological sex classification in the UKB dataset. For ADHD diagnosis and biological sex classification in the ABCD dataset, and fluid intelligence regression in the UKB dataset, the Quantum Time-series Transformer demonstrated comparable performance to classical models, with the classical BolT model achieving slightly better results in these tasks (Table \ref{FullSample}).

\begin{table*}[!ht]
    \centering
    \caption{Comparison of Test Performance from Full Samples}
    \begin{tabular}{cccccc}
    \toprule
        ~ & ~ & ABCD & ~ & \multicolumn{2}{c}{UKB} \\ 
        \cmidrule(r){2-4} \cmidrule(r){5-6}
        ~ & Sex (AUROC) & ADHD (AUROC) & Fluid Intelligence (MAE) & Sex (AUROC) & Fluid Intelligence (MAE) \\ 
        \midrule
        Sample Size & N = 9,363 & N = 4,550 & N = 5,472 & N = 40,792 & N = 21,495 \\
        \midrule
        Vanilla Transformer & 0.5328 ± 0.0192 & 0.5613 ± 0.0267 & 0.7717 ± 0.0021 & 0.9485 ± 0.0004 & 0.7897 ± 0.0014 \\ 
        BNT  & 0.7253 ± 0.0223 & 0.5503 ± 0.0026 & 0.7825 ± 0.0075 & 0.9437 ± 0.0001 & 0.7910 ± 0.0011 \\ 
        BolT  & \textbf{0.8185 ± 0.0067} & \textbf{0.6290 ± 0.0097} & 0.7704 ± 0.0019 & 0.9493 ± 0.0004 & \textbf{0.7857 ± 0.0008} \\ 
        \textbf{QuantumTSTransformer}  & 0.8182 ± 0.0037 & 0.6131 ± 0.0066 & \textbf{0.7683 ± 0.0019} & \textbf{0.9557 ± 0.0004} & 0.7879 ± 0.0011 \\ 
        \bottomrule
    \end{tabular}
    \label{FullSample}
\end{table*}

\begin{table*}[!ht]
    \centering
    \caption{Comparison of Test Performance from Small Samples}
    \begin{tabular}{cccccc}
    \toprule
        ~ & ~ & ABCD & ~ & \multicolumn{2}{c}{UKB} \\ 
        \cmidrule(r){2-4} \cmidrule(r){5-6}
        ~ & Sex (AUROC) & ADHD (AUROC) & Fluid Intelligence (MAE) & Sex (AUROC) & Fluid Intelligence (MAE) \\ 
        \midrule
        Sample Size & N = 100 & N = 100 & N = 100 & N = 100 & N = 100 \\
        \midrule
        Vanilla Transformer & 0.5167 ± 0.1146 & 0.4976 ± 0.0728 & 0.6452 ± 0.0534 & 0.8393 ± 0.1014 & 1.0344 ± 0.0818 \\ 
        BNT  & 0.4742 ± 0.0954 & 0.5262 ± 0.0205 & 0.7050 ± 0.0356 & 0.6310 ± 0.1248 & 1.3304 ± 0.1918 \\ 
        BolT  & 0.4141 ± 0.0430 & 0.5206 ± 0.0600 & 0.7329 ± 0.0626 & 0.8988 ± 0.1012 & 1.0969 ± 0.1036 \\ 
        \textbf{QuantumTSTransformer}  & \textbf{0.5205 ± 0.0295} & \textbf{0.6963 ± 0.1123} & \textbf{0.6095 ± 0.0616} & \textbf{0.9405 ± 0.0315} & \textbf{0.8988 ± 0.0462} \\ 
        \bottomrule
    \end{tabular}
    \label{SmallSample}
\end{table*}

When the sample size was reduced to N = 100 (i.e., 70 for training, 15 for validation, and 15 for testing), the Quantum Time-series Transformer outperformed all classical counterparts across both the ABCD and UKB datasets. The quantum model exhibited higher AUROC in biological sex and ADHD classification and lower MAE in fluid intelligence regression compared to all classical transformers (Table \ref{SmallSample}).

\begin{figure*}[!ht]
    \centering
    \begin{subfigure}[ht]{0.5\textwidth}
        \centering
        \includegraphics[height=2.4in]{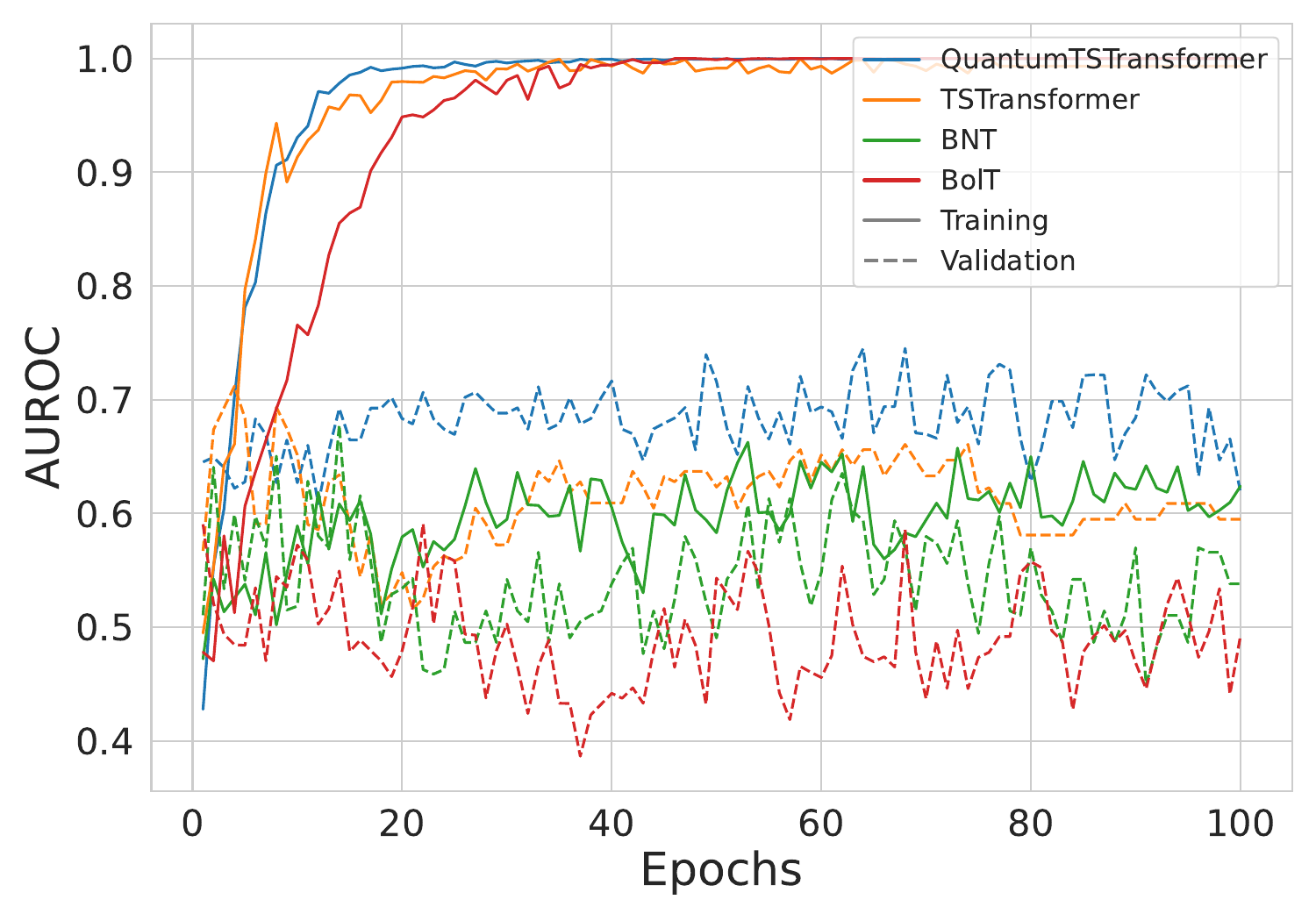}
        \caption{ABCD ADHD Diagnosis Classification (N=100)}
    \end{subfigure}%
    \hfill    
    \begin{subfigure}[ht]{0.5\textwidth}
        \centering
        \includegraphics[height=2.4in]{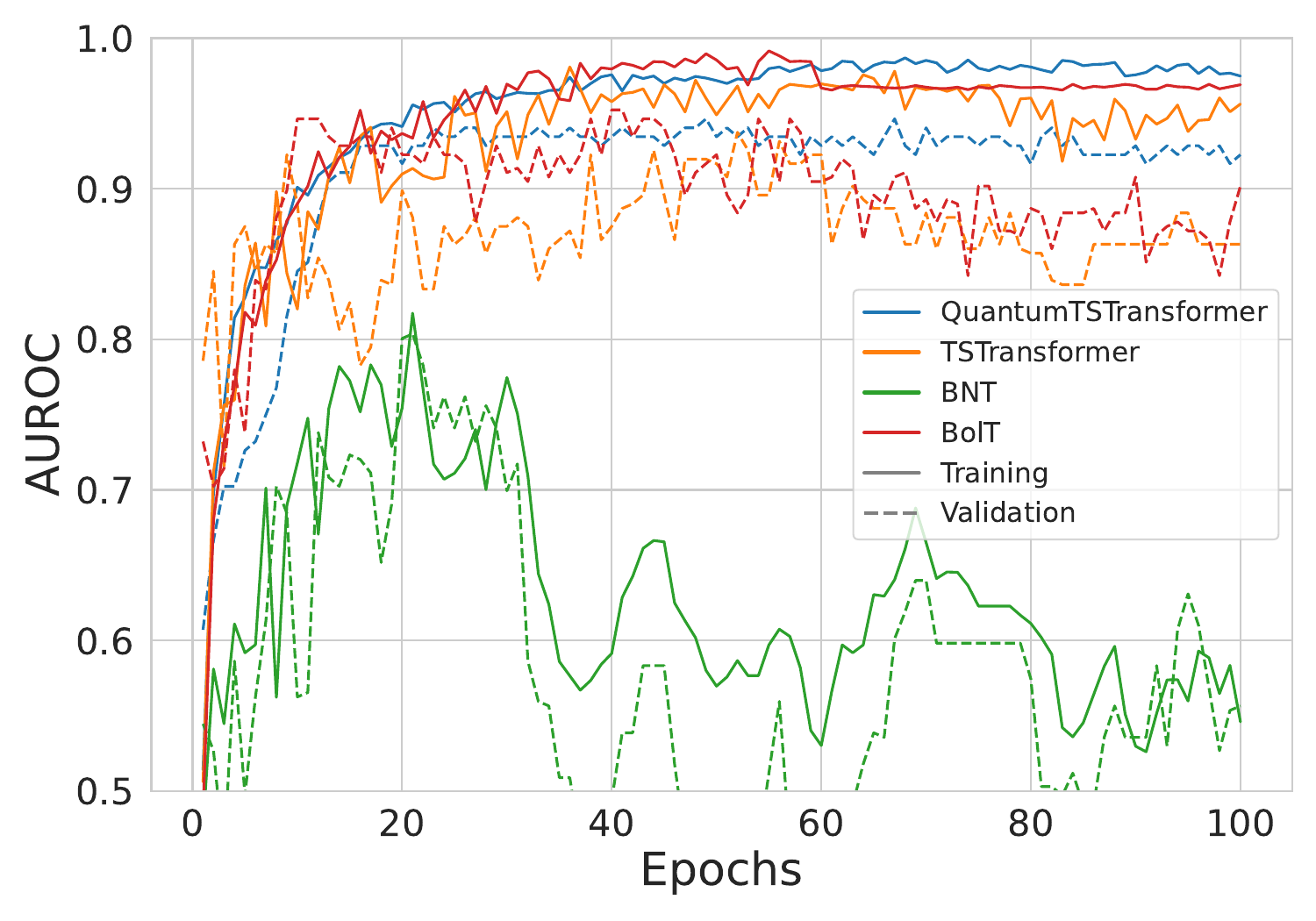}
        \caption{UKB Sex Classification (N=100)}
    \end{subfigure}
    \caption{Results from ABCD ADHD diagnosis and UKB biological sex classication in small sample sizes}    
    \label{Fig_Performance}
\end{figure*}

Furthermore, the Quantum Time-series Transformer showed faster convergence compared to the classical models. As shown in Figure \ref{Fig_Performance}, the Quantum Time-series Transformer reached a higher validation AUROC more quickly than any classical transformer. Additionally, the Quantum Time-series Transformer exhibited better generalizability, as evidenced by its lower generalization error—the difference between training and validation/test performance—compared to the classical models.

Notably, Quantum Time-series Transformer achieved such performance with significantly fewer trainable parameters than any other classical transformers that we assessed. While classical transformers required at least 1.68 million parameters for training, the Quantum Time-series Transformer was composed of only 22 thousand parameters (Table \ref{NumberofParameters}). This means that the quantum model had 75 to 500 times fewer number of parameters than the classical counterparts.

\begin{table}[!ht]
    \centering
    \caption{Comparison of Number of Trainable Parameters}
    \begin{tabular}{cc}
    \toprule
        ~ & Number of Parameters \\ 
        \midrule
        Vanilla Transformer & 1.68M \\ 
        BNT  & 11.2M \\ 
        BolT  & 1.68M \\ 
        QuantumTSTransformer  & 22K \\ 
        \bottomrule
    \end{tabular}
    \label{NumberofParameters}
\end{table}

\begin{figure*}[h]
    \centering
    \includegraphics[width = 0.9\textwidth]{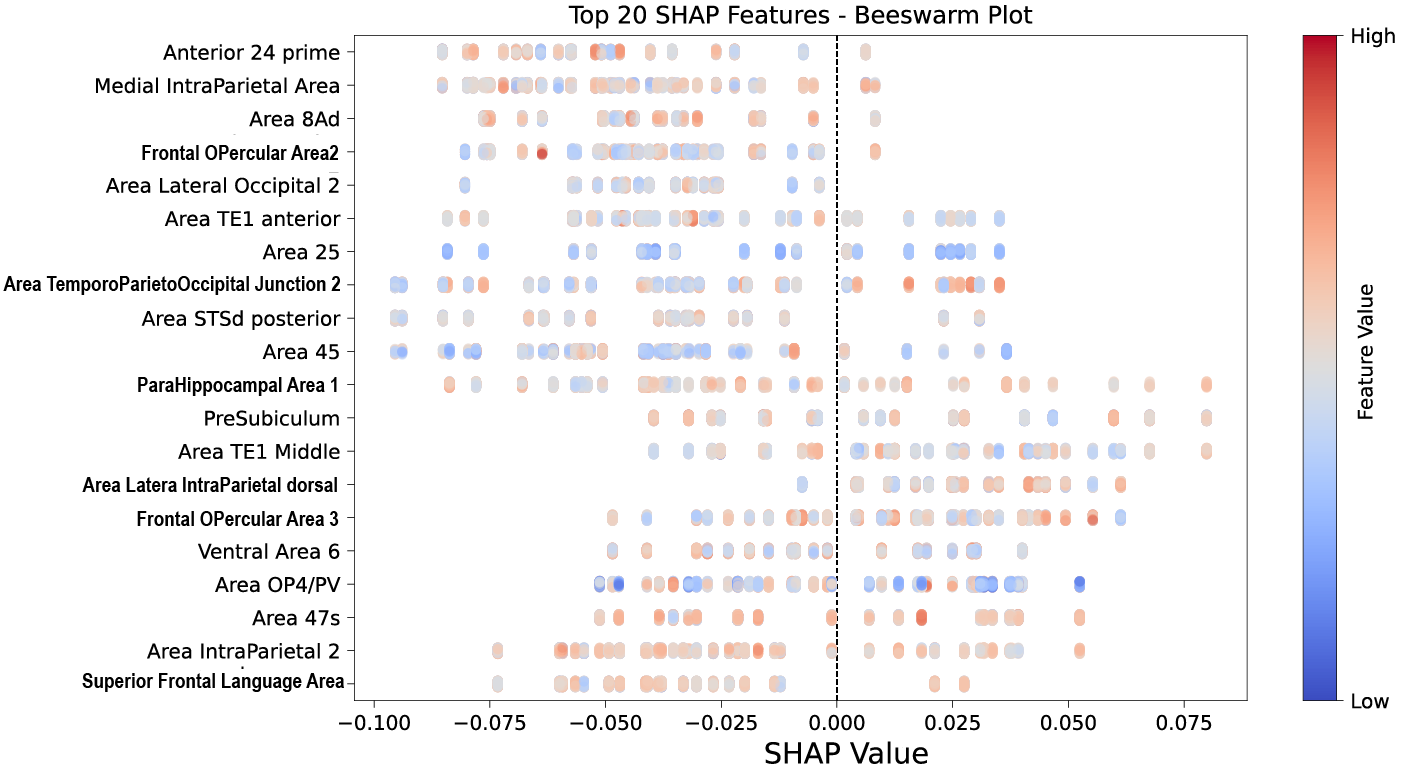}
    \caption{Model Interpretation using SHAP Beeswarm Plot for ADHD Diagnosis Classification} 
    \label{Fig_SHAP}
\end{figure*}

To interpret the resting-state fMRI results of the Quantum Time-series Transformer and derive scientific implications, we obtained SHAP beeswarm plot illustrating the top 20 most predictive brain regions for ADHD diagnosis (Fig. \ref{Fig_SHAP}). The distribution of SHAP values reveals the directionality and magnitude of each region’s contribution to the model’s classification decision. Notably, higher SHAP values indicate greater positive influence on ADHD prediction, while lower SHAP values indicate stronger negative contributions.

Notable brain regions identified by the Quantum Time-series Transformer include prefrontal and opercular areas (e.g., Anterior 24, Area 8Ad, Frontal Opercular Areas 2 \& 3, Area 45, Area 47s), intra-parietal regions (e.g., Medial IntraParietal, Lateral IntraParietal, Area IntraParietal 2), and limbic structures (e.g., ParaHippocampal Area 1, Area 25, PreSubiculum). These regions are strongly associated with executive function, attention control, motor inhibition, and reward processing—key cognitive processes implicated in ADHD pathology \cite{Sudre, Lampert, Hauser}. Additionally, regions involved in visual and multisensory integration (e.g., Area Lateral Occipital 2, TemporoParietoOccipital Junction 2, Superior Frontal Language Area) suggest alterations in sensory processing and cognitive control, consistent with ADHD-related deficits, \cite{Bush, Sudre}.

The Quantum Time-series Transformer’s ability to capture spatio-temporal dependencies in fMRI time series enables a more nuanced understanding of brain dynamics relevant to ADHD, as shown by the SHAP contributions. This supports the potential of quantum-enhanced deep learning models in psychiatric neuroimaging, particularly in identifying neural biomarkers with greater interpretability and clinical relevance.

\section{Conclusion}
This study proposed and validated the Quantum Time-series Transformer, a novel quantum-enhanced deep learning algorithm specifically crafted for resting-state fMRI data analysis. By leveraging quantum computational primitives such as LCU and QSVT, Quantum Time-series Transformer effectively captured complex spatio-temporal dependencies in realistic, high-dimensional neuroimaging data. Our comparative evaluations against state-of-the-art classical transformer models (e.g., BolT and BNT) on ADHD diagnosis, biological sex classification, and fluid intelligence prediction consistently demonstrated that the quantum approach could match or surpass classical baselines.

A notable strength of Quantum Time-series Transformer is its ability to achieve robust performance with fewer parameters and smaller training samples \cite{Caro2022}. In typical neuroimaging research—where collecting large, well-curated datasets is challenging—high-parameter classical transformers often risk overfitting and become computationally prohibitive \cite{Marek2022}. The Quantum Time-series Transformer’s capacity to generalize under data-sparse conditions, as evidenced by its success in restricted-sample scenarios (e.g., N=100), addresses a critical bottleneck for applying deep learning to many neuroscience and psychiatry studies.


Furthermore, while many QML studies have demonstrated quantum advantages primarily in small-scale toy datasets, their practical utility for real-world high-dimensional tasks remains limited \cite{Schuld}. Recent studies have expressed skepticism regarding the practicality of QML approaches, especially those based on VQCs \cite{Qian, Rani}. Addressing these concerns, our study demonstrated that the Quantum Time-series Transformer could achieve strong performance on realistic, high-dimensional spatio-temporal brain data. Importantly, we deliberately avoided a restrictive experimental setup biased toward quantum algorithms, opting instead for a realistic scenario common in neuroscience research involving resting-state fMRI data.

Beyond performance metrics, our interpretability analysis using SHAP highlighted Quantum Time-series Transformer’s ability to identify clinically meaningful neural biomarkers of ADHD. Key brain regions involved in executive control, attentional processing, and sensory integration emerged as salient features, confirming findings from established neuropsychiatric literature. Notably, the Quantum Time-series Transformer framework integrates SHAP values directly into quantum-specific operations (LCU and QSVT), adding novel insights into the quantum model’s decision mechanisms. This result offers both theoretical significance—by illustrating how quantum-temporal modeling can be rendered more transparent—and practical value for clinical applications, including biomarker discovery and individualized diagnostic profiling.


Unlike previous work that primarily addressed gate-level interpretations of quantum circuits \cite{Heese} or local interpretability of simpler VQCs \cite{Pira}, our approach uniquely integrates SHAP interpretability directly into quantum attention mechanisms—specifically LCU and QSVT.


Despite these promising outcomes, several limitations warrant discussion. First, our model employed classical dimensionality reduction methods prior to quantum circuit analysis, which may lead to information loss regarding the underlying high-dimensional neural data. Future research could address this by developing quantum algorithms capable of processing complex, high-dimensional data directly, eliminating the need for classical dimensionality reduction. Second, although our analysis leveraged large-scale neuroimaging datasets (ABCD and UKB), the generalizability of these findings across more diverse clinical populations and scanning conditions remains to be validated \cite{Henrich}. Finally, although we provided meaningful interpretability through SHAP analysis, future studies could conduct more in-depth clinical and neuroscientific analyses using alternative interpretability methods specifically tailored for quantum-enhanced models, such as Quantum Grad-CAM \cite{Lin}.

In conclusion, this work illustrates how quantum-enhanced transformers can open new avenues in computational neuroscience, offering a compact, interpretable approach that remains effective under constraints of limited data or computational resources. By bridging quantum computing principles with clinically meaningful biomarkers of brain function, the Quantum Time-series Transformer demonstrates a promising paradigm for scalable and transparent neuroimaging analyses. We anticipate that future refinements—particularly around quantum hardware implementation and methodological rigor—will further bolster the Quantum Time-series Transformer’s potential to transform the landscape of neuropsychiatric research and precision medicine.

\section*{\textbf{Acknowledgment}}
The authors extend their gratitude to the members of the Connectome Lab for their invaluable support and critical contributions to the development of the Quantum Time-series Transformer. This work was supported by the National Research Foundation of Korea (NRF) grant funded by the Korea government (MSIT) (No. 2021R1C1C1006503, RS-2023-00266787, RS-2023-00265406, RS-2024-00421268, RS-2024-00342301, RS-2024-00435727, NRF-2021M3E5D2A01022515), by Creative-Pioneering Researchers Program through Seoul National University (No. 200-20240057, 200-20240135), by Semi-Supervised Learning Research Grant by SAMSUNG (No.A0342-20220009), by Identify the network of brain preparation steps for concentration Research Grant by LooxidLabs (No.339-20230001), by Institute of Information \& Communications Technology Planning \& Evaluation (IITP) grant funded by the Korea government (MSIT) [NO.RS-2021-II211343, Artificial Intelligence Graduate School Program (Seoul National University)] by the MSIT (Ministry of Science, ICT), Korea, under the Global Research Support Program in the Digital Field program (RS-2024-00421268) supervised by the IITP (Institute for Information \& Communications Technology Planning \& Evaluation), by the National Supercomputing Center with supercomputing resources including technical support (KSC-2023-CRE-0568) and by the Ministry of Education of the Republic of Korea and the National Research Foundation of Korea (NRF-2021S1A3A2A02090597), by the Korea Health Industry Development Institute (KHIDI), and by the Ministry of Health and Welfare, Republic of Korea (HR22C1605), by Artificial intelligence industrial convergence cluster development project funded by the Ministry of Science and ICT (MSIT, Korea) \& Gwangju Metropolitan City, by KBRI basic research program  through  Korea  Brain  Research  Institute funded by Ministry of Science and ICT(25-BR-05-01), and by the education and training program of the Quantum Information Research Support Center, funded through the National Research Foundation of Korea (NRF) by the Ministry of Science and ICT (MSIT) of the Korean government (No.2021M3H3A1036573). This work was in part supported by the U.S. Department of Energy  (DOE), Office of Science (SC), Advanced Scientific Computing Research program under award DE-SC-0012704.

\clearpage

\bibliographystyle{ieeetr}
\bibliography{references}

\end{document}